\documentclass[12pt]{article}

\usepackage{amsmath}
\usepackage{amsfonts}
\usepackage{amssymb}

\usepackage[dvips]{epsfig}
\usepackage[dvips]{graphics}

\renewcommand{\baselinestretch}{1.25}

\begin{document}

\title{\textbf{Constraints on the Detectability of Cosmic Topology from
Observational Uncertainties}}
\author{B. Mota$^{1}$\thanks{brunom@cbpf.br}, \
M.J. Rebou\c{c}as$^{1}$\thanks{reboucas@cbpf.br} \ and \ R.
Tavakol$^{2}$\thanks{r.tavakol@qmw.ac.uk} \\
\\
$^{1}$ Centro Brasileiro de Pesquisas F\'\i sicas, \\ Rua Dr.
Xavier Sigaud 150
\\ 22290-180 Rio de Janeiro -- RJ, Brazil\\
\\
$^{2}$ Astronomy Unit, School of Mathematical Sciences, \\
Queen Mary, University of London, \\
Mile End Road, London E1 4NS, UK }
\date{\today }
\maketitle

\renewcommand{\baselinestretch}{0.8}
\begin{abstract}\noindent
Recent observational results suggest that our universe is nearly
flat and well modelled within a $\Lambda$CDM framework. The
observed values of $\Omega_{m}$ and $\Omega_{\Lambda}$ inevitably
involve uncertainties. Motivated by this, we make a systematic
study of the necessary and sufficient conditions for undetectability
as well as detectability (in  principle) of cosmic topology (using
pattern repetition) in presence of such uncertainties. We do this by
developing two complementary methods to determine detectability
for nearly flat universes. Using the first method we derive
analytical conditions for undetectability for infinite
redshift, the accuracy of which is then confirmed by the
second method. Estimates based on WMAP data together with other
measurements of the density parameters are used to illustrate
both methods, which are shown to provide very similar results
for high redshifts.
\end{abstract}

\newpage
\renewcommand{\baselinestretch}{1.25}
\section{Introduction}
\label{intro}

An important feature of general relativity is that it is a local
metrical theory, and therefore the corresponding Einstein field
equations do not fix the global topology of space-time. This
freedom has fueled a great deal of interest in the possibility
that the universe may possess compact spatial sections with a
non-trivial topology (see for example~\cite{CosmicTop,Revs}).
Whatever the nature of cosmic topology may turn out to be, the
issue of its detectability is of fundamental importance.

Motivated by recent observational results, a study was recently
made of the question of detectability of the cosmic topology in
nearly flat universes where the ratio of the current total density
to the critical density of the universe, $\Omega_{0}$, is very
close to one. It was demonstrated that as $\Omega_{0}\rightarrow1$
increasing families of possible topologies become undetectable by
methods based on image (or pattern) repetitions
(see~\cite{grt2001a}~--~\cite{NEWweeks2}).

Measurements of the density parameters unavoidably involve
observational uncertainties, and therefore any study of the
detectability of the cosmic topology should take such
uncertainties into account. This is particularly crucial for
nearly flat universes, which are favoured by the current
observations.

In this paper we study the sensitivity of the detectability of
cosmic topology to the uncertainties in the density parameters.
We present two complementary methods for deciding the
detectability of cosmic topology in terms of the density parameters
within the uncertainty region, for any given survey depth. The first
method provides sufficient (but not necessary) conditions for
undetectability of cosmic topology. The second method provides
necessary (but not sufficient) conditions for undetectabilty.
The converses of the latter conditions also give sufficient
(but not necessary) conditions for detectability in principle
of cosmic topology.
Using the former method in the limiting case $z\rightarrow\infty$
we derive an exact closed form, which expresses the sufficient
conditions for undetectability of cosmic topology of nearly
flat universes.

Both methods were devised to be suitable where the values of
density parameters include uncertainties, which lie in a region
in the neighbourhood of $\Omega_{0}=1$, and are shown to be accurate
for high redshifts. Numerical criteria for both undetectability and
detectability in principle (collectively denoted in what follows
by (un)de\-tect\-abil\-i\-ty to be succinct) are also presented.

The structure of the paper is as follows. In section~\ref{prelim}
we give an account of the cosmological models employed, present
a brief discussion of a topological indicator we will use, and make
a brief analytical study of the question of detectability of
cosmic topology of nearly flat universes.
In section~\ref{criteria} we develop two complementary methods for
deciding the detectability of cosmic topology taking into account
the uncertainties in the density parameters, for any given survey
depth. In section~\ref{Examples} we present a number of concrete
examples, and discuss their connection with some results in the
literature.
We construct tables for specific topologies which provide
support for the assertion that the closed form expression that ensures
sufficient conditions for undetectability is very accurate for
deciding the undetectability of cosmic topology of nearly flat
universes.
Finally, section~\ref{Conclusion} contains a discussion of our
main results and conclusions.

\section{Preliminaries}
\label{prelim}

In the context of standard cosmology, the universe is modelled by
a $4$-manifold $\mathcal{M} = \mathcal{R}\times M$, with a locally
isotropic and homogeneous Robertson-Walker (RW) metric
\begin{equation} \label{FLRW1}
ds^{2}=-c^{2}dt^{2}+R^{2}(t)\left[d\chi^{2}+f^{2}(\chi)(d\theta^{2}
+\sin ^{2}\theta\,d\phi^{2})\right] \;,
\end{equation}
where $t$ is a cosmic time, $f(\chi)=\chi\,$, $\sin\chi$, or
$\sinh\chi\,$ depending on the sign of the constant spatial
curvature ($k=0,\pm1$), and $R(t)$ is the scale factor. For
non-flat models ($k\neq0$), the scale factor is identified with
the curvature radius of the spatial section of the universe at
time $t$. Usually the $3$-space $M$ is taken to be
simply-connected; namely Euclidean $E^{3}$, spherical $S^{3}$, or
hyperbolic $H^{3}$ spaces. In general, however, the $3$-space may
take the form of an infinite set of other possible quotient
(multiply connected) manifolds $M=\widetilde{M}/\Gamma$, where
$\Gamma$ is a discrete group of freely acting isometries of the
covering space $\widetilde{M}$.%
\footnote{In this article, in line with
the usage in the literature, by topology of the universe we
mean the topology of the space-like section $M$ of the space-time
manifold $\mathcal{M}$.}

Recent observations have provided important information concerning
the nature of the energy content of the universe. In particular,
recent measurements by WMAP~\cite{WMAP} of the position of the
first acoustic peak in the angular power spectrum of cosmic
microwave background radiation anisotropies (CMBR), which refine
and to a great extent corroborate previous data%
~\cite{QMAP-MAT-TOCO}~--~\cite{MAXIMA}), seem to suggest that
the universe is flat or nearly so ($\Omega_0 \sim 1$).

There is also ample evidence from observations, including the CMBR
power spectrum, galaxy clustering statistics~\cite{GlxyClstrStat},
peculiar velocities~\cite{PecVeloc} and the baryon mass fraction
in clusters of galaxies~\cite{BrynFrac,StrFormBound} that
the density of the clumped (including baryonic and dark) matter in
the universe is substantially lower, being of the order of $0.3$
of the critical value. Furthermore, observations
of high redshift Type Ia Supernovae~\cite{perl} seem to suggest
that the universe is presently undergoing accelerated expansion.

One way of reconciling these diverse set of observations is to
postulate that a substantial proportion of the energy density of
the universe is in the form of a dark component which is smooth on
cosmological scales and possesses a negative pressure. An
important candidate for this is the cosmological constant.

In the present work we therefore assume that the current matter content
of the universe is well approximated by dust (of density $\rho_{m}$) plus
a cosmological constant $\Lambda$, with associated fractional densities
$\Omega_{m}=8 \pi G\rho_{m}\,/\,(3H^2)$ and $\Omega_{\Lambda}%
\equiv \Lambda\,c^2/\,(3H^2)$, with
$\Omega_{0}=\Omega_{m}+\Omega _{\Lambda}$. In this setting, for
non-flat models, the redshift-(comoving)-distance relation in units of
curvature radius takes the form
\begin{equation} \label{chi(z)}
\chi(z)=\sqrt{|1-\Omega_{0}|}\int_{0}^{z}\left[ (1+x)^{3}\Omega_{m0}%
+\Omega_{\Lambda0}-(1+x)^{2}(\Omega_{0}-1)\right]^{-\frac{1}{2}}dx\;,
\end{equation}
where the redshift $z$ measures the depth of the survey, and the
subscript '$0$' refers to present values of the density
parameters. The horizon radius $d_{hor}$ is then def\/ined by~%
(\ref{chi(z)}) for $z = \infty\,$.
For simplicity, on the right hand side of~(\ref{chi(z)})
and in many places in the remainder of this article, we have left
implicit the dependence of the function $\chi$ on the density components.

An important feature of the expression~(\ref{chi(z)}) is that it is
very sensitive to changes in the quantity $1-\Omega_{0}$ near the
flat line, falling rapidly to zero as $\Omega_{0} \rightarrow
1$ (as will become quantitatively evident below). This limit plays
a crucial role in the detectability of any non-trivial
topology~\cite{grt2001a}~--~\cite{grt2002}.

To proceed we recall some topological background that we shall use
below. To begin with we note that the classes of topologies
allowed for spherical, flat and hyperbolic $3$-manifolds are
qualitatively different, and any analysis must deal with each
family separately. Furthermore, in three dimensions there are
complete classifications for flat and spherical topologies, but
not for the hyperbolic ones. Also, geometrical quantities of
hyperbolic and spherical manifolds, expressed in terms
of the curvature radius, are topological invariants. For flat
manifolds, however, we have no such natural unit of length, and
they are not rigid. They should therefore be dealt with separately
(see~\cite{gr2002} for a study of detectability in such cases).
Here we shall confine ourselves to non-flat cases only.

Compact orientable hyperbolic 3-manifolds can be constructed by the
so called Dehn surgery procedure, denoted by two coprime winding
numbers, $(n_1,n_2)$, applied to a cusped seed manifold. For
example, in the case of the manifold $m003(-3,1)$ (the smallest
known orientable hyperbolic manifold), $m003$ is the seed cusped
manifold, and $-3$ and $1$ are the winding numbers. Using the
software SnapPea~\cite{SnapPea}, Hodgson and Weeks~\cite{HW}
compiled a census with 11031 of such manifolds, ordered by
volume.

In order to study the (possibly non-trivial) topology of the
spatial sections $M$ of the universe, we need a topological
invariant length that could be put into correspondence with
depth of surveys. We shall employ the so-called injectivity radius
$r_{inj}$, the radius of the smallest sphere 'inscribable' in $M$,
which is defined as half the length $\ell_{M}$ of the smallest
closed geodesics (for details see~\cite{grt2001a}),
\begin{equation}
r_{inj}=\frac{\ell_{M}}{2}\;. \label{rinj}
\end{equation}
The indicator $T_{inj}$, i.e. the ratio of the injectivity radius
to the depth $\chi(z_{obs})$ of a given astro-cosmological survey
up to a maximum redshift $z=z_{obs}$, is then defined as
\begin{equation}
T_{inj}=\frac{r_{inj}}{\chi(z_{obs})}\;.\label{r_inj}%
\end{equation}
In a universe with $T_{inj}>1$ there would be no observed multiple
images of objects up to the maximum survey depth, and therefore
the cosmic topology would not be detectable observationally by any
observer looking for patterns repetition. Similarly for universes
with $T_{inj}<1$ the topology is in principle observationally
detectable through pattern repetitions, at least for some
observers.

An important issue (and the main goal for this work) is to
determine the which manifolds (topologies) are undetectable, and
which are detectable (in principle), for given values of the density
parameters ($\Omega_{m0}$, $\Omega_{\Lambda0}$) subject to
observational uncertainties. If $\Omega_{m0}$ and
$\Omega_{\Lambda0}$ were known precisely, it would be a simple
matter to calculate $\chi(z_{obs})$ and then obtain $T_{inj}$ for
any non-flat universe (with the associated $r_{inj}$). The density
parameters, however, inevitably involve observational uncertainties,
with an associated uncertainty region in the parameter plane.
The most recent estimates~\cite{WMAP} specify this region to be
$\Omega_{0}\in [0.99,1.05]$ and $\Omega_{\Lambda }\in [0.69,0.79]$
with a $2\sigma$ confidence, straddling the flat line
where $\chi(z)=0$. So, for any $3$-manifold there is a set of values
for $\Omega_{m0}$ and $\Omega_{\Lambda0}$ within the uncertainty
region for which the corresponding topology is undetectable.
The (un)detectability issue has been investigated, for some specific
values of the density parameters in~\cite{grt2001a}~--~\cite{grt2002}
and~\cite{NEWweeks}.  Here we systematically
extend those results by developing methods that for each
manifold $M$ separates the uncertainty region into undetectable
and  detectable (in principle) sub-regions, for any given survey
depth $z_{obs}$.

\section{Conditions for undetectability and detectability in principle}
\label{criteria}

\begin{figure}[!tbh]
\centerline{\def\epsfsize#1#2{0.7#1}\epsffile{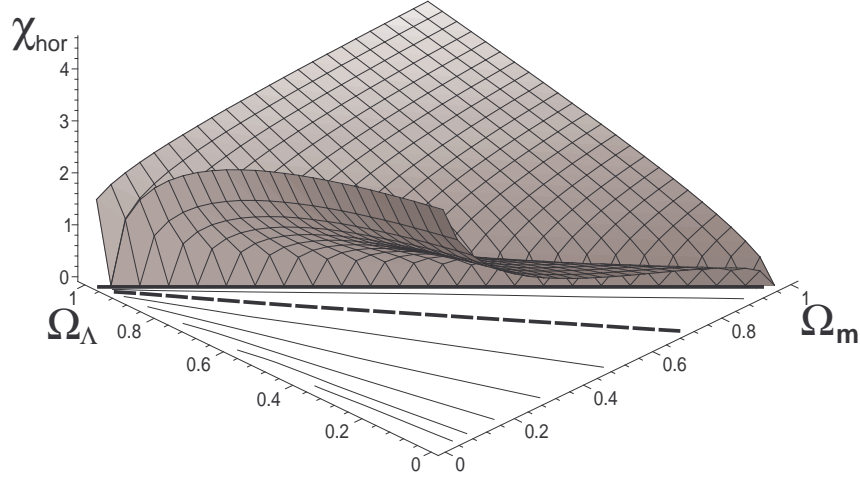}}
\caption{The behaviour of the horizon radius $\chi_{hor}$
in units of curvature radius as a function of the density
parameters $\Omega_{\Lambda}$ and $\Omega_m\,$. Clearly a
similar behaviour holds for $\chi_{obs}$ for any fixed $z_{obs}\,$.
The curves in the parametric plane are contour curves defined by
$\chi(\Omega_{m0},\Omega_{\Lambda0},z_{obs}) = r_{inj}^{\:M}\,$
for a given manifold with $r_{inj}=r_{inj}^{\:M}\,$, and a fixed
survey depth $z_{obs}\,$.   \label{BirdContFig} }
\end{figure}

To motivate our methods we begin by noting that a first  estimate
of the constraints on detectability of cosmic topology can be obtained
from the horizon radius function
$\chi_{hor}(\Omega_{m0},\Omega_{\Lambda0},z)\,$ given by~(\ref{chi(z)})
for $z = \infty\,$, in the neighbourhood of the  flat line
$\Omega_0= \Omega_{m0}+\Omega_{\Lambda 0}=1$ favoured by recent
diverse set of observations.
Figure~\ref{BirdContFig} shows the behaviour of $\chi_{hor}$ as
a function of $\Omega_{m0}$ and $\Omega_{\Lambda 0}$.
The curves in the parametric plane are contour curves defined by
$\chi(\Omega_{m0},\Omega_{\Lambda0},z_{obs}) = r_{inj}^{\:M}\,$
for a given manifold with $r_{inj}=r_{inj}^{\:M}\,$, and a fixed
survey depth $z_{obs}\,$.
Since in parametric plane the flat universes are characterized by
the straight line $\Omega_0=1$, this figure makes clear that as
$|\Omega_0 -1| \to 0$, $\chi_{hor} \to 0$, hence showing that,
for a given manifold $M$ with injectivity radius $r_{inj}^{\:M}\,$
there are values of $\Omega_{\Lambda 0}$ and $\Omega_{m0}$ for
which the topology of the universe is either undetectable
($T_{inj}>1\,$, i.e. $\,\,r_{inj}^{\:M}\, > \chi_{hor}\,$) or
detectable in principle ($r_{inj}^{\:M}\, < \chi_{hor}\,$).

Such scheme has been employed for specific values of the density
parameters (see, e.g., ~\cite{grt2001a}~--~\cite{grt2002}). Here we
extend this approach so as to include the whole observational
uncertainty region $\,\mathcal{U}\,$, which is determined by
a diverse set of observations (CMB, SNIa, lensing, large scale
structure observations). Furthermore, in the limiting case
$z\rightarrow\infty$ we shall derive an exact closed form,
which expresses the sufficient conditions for undetectability
of cosmic topology of nearly flat universes, no matter how
complicated is the spatial topology of $M$.

\begin{figure}[!tbh]
\centerline{\def\epsfsize#1#2{0.7#1}\epsffile{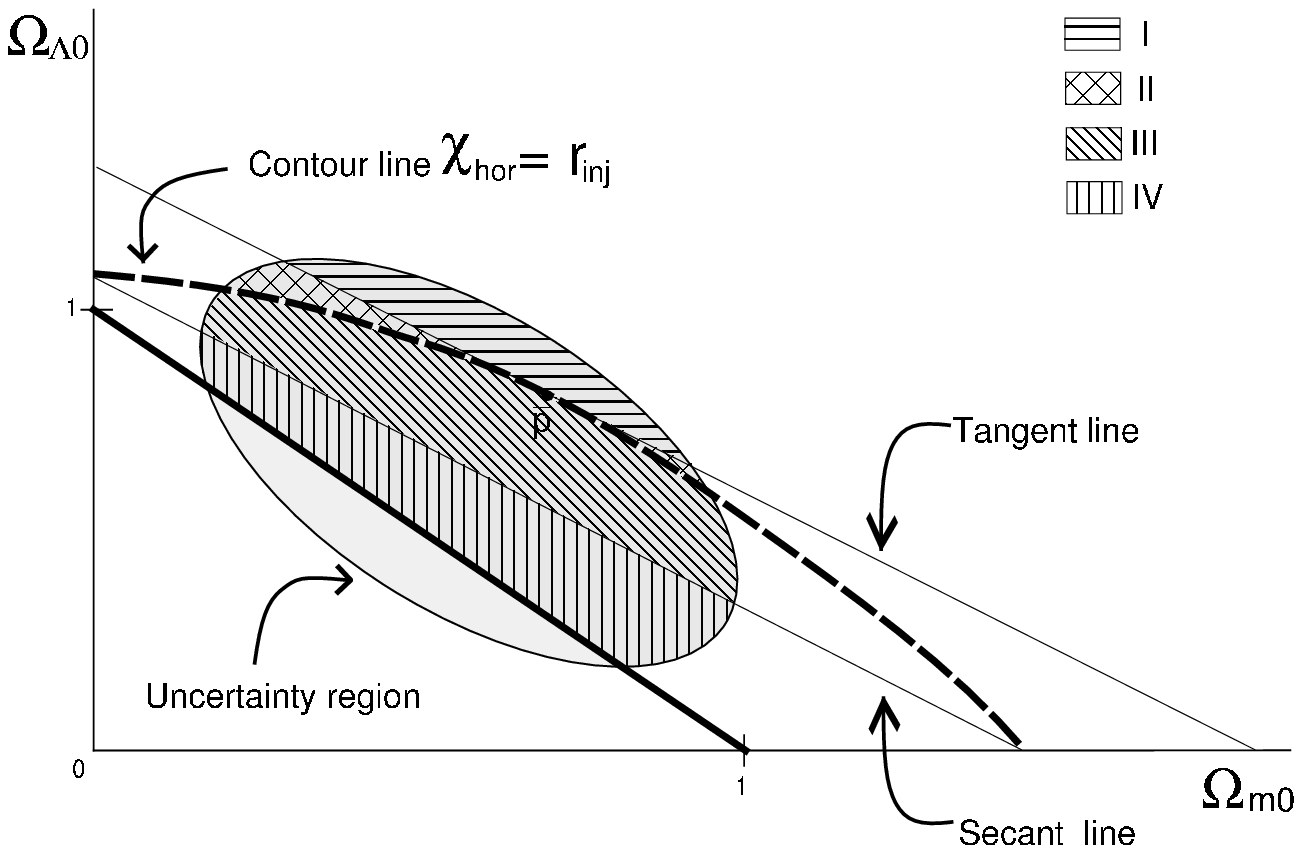}}
\caption{A schematic representation of the secant line (SL) and
tangent line (TL) methods. The convexity of the contour curve for
$\Omega_{0}>1$ can be proven analytically, as discussed below.
The topology is shown to be detectable in principle in region I by
the TL method, and undetectable in region IV by the SL method.
Regions II and III are not discriminated by either methods. But
using just the contour curve one has that in region II the topology
is detectable in principle, whereas in region III it is undetectable.
The sub-region of the uncertainty region $\,\mathcal{U}\,$ between
the secant and the tangent lines (regions II and III) is very small
for manifolds whose contour curves intersect the uncertainty region.
\label{SchemFig} }
\end{figure}

From now on we shall focus our attention in the parametric plane
$\Omega_{\Lambda0}$~--~$\Omega_{m0}$.
To better understand undetectability regarding this plane consider
a manifold $M$ with $r_{inj}=r_{inj}^{\:M}\,$. For a fixed survey
depth $z=z_{obs}$ a contour curve is defined by
$\chi(\Omega_{m0},\Omega_{\Lambda0},z_{obs}) = r_{inj}^{\:M}\,$.
This curve defines two sub-regions in the parametric plane: one
between the flat line and the contour curve where the topology
of a universe with space section $M$ is undetectable for a survey
with depth $z_{obs}$, and another beyond the contour line where it
is detectable in principle for the same survey depth
(see Fig.~\ref{SchemFig}). If the contour curve does not intersect
the current uncertainty region $\mathcal{U}$ the topology of $M$ is
undetectable for the values of the density parameters in $\mathcal{U}$.
If it does, a criterion is needed to determine the set of values of
the density parameters in $\mathcal{U}$ for which the topology
is undetectable.

We shall use two different methods to obtain linear approximations
of the contour curves.
The first approximation method (called secant method) provides
sufficient (but not necessary) conditions for undetectabilty of
cosmic topology. As such, these conditions do not apply to the
whole region where the topology of $M$ is undetectable, since
there is a sub-region of $\,\mathcal{U}\,$ above the secant line
for which the topology of $M$ is still undetectable
(see Fig.~\ref{SchemFig}). To ensure that the secant line approximation
method is an efficient method to decide the detectability in practice,
we need to show that this sub-region is small. We do this by employing
a second linear approximation method (which we refer to as the tangent
method) that provides necessary (but not sufficient) conditions for
undetectabilty. Since both methods rely on linear approximation, the
conditions provided by them will be given in terms of linear
inequalities in the density parameters.

The secant line method simply states that a universe with space-like
section $M$ has an undetectable topology if
\begin{equation}  \label{crite1}
\alpha\;\Omega_{m0}+\beta\;\Omega_{\Lambda0}\left\{
\begin{array} [c]{c}%
>1\;,  \quad \mathrm{for} \quad  \Omega_{0} < 1 \;,\\
<1\;,  \quad  \mathrm{for} \quad \Omega_{0} > 1 \;,
\end{array}
\right.
\end{equation}
where the coefficients $\alpha$, $\beta$ are ``normalized'' constants
fixed by the topology (via $r_{inj}^{\:M}$) and the depth of the
survey $z_{obs}$.

Similarly, the tangent method states that the necessary
conditions for $M$ to have undetectable topology are
\begin{equation} \label{crite2}
\mu \;\Omega_{m0} +
\nu\;\Omega_{\Lambda0}\left\{
\begin{array} [c]{c}%
<1\;,  \quad \mathrm{for} \quad\Omega_{0} >1 \;, \\
>1\;,  \quad
\mathrm{for} \quad \Omega_{0} < 1 \;,
\end{array}
\right.
\end{equation}
where again the coefficients $\mu$, $\nu$ are ``normalized'' constants
fixed by the topology (via $r_{inj}^{\:M}$) and the depth of the
survey $z_{obs}$. We note that the converses of conditions%
~(\ref{crite2}) provide sufficient (but not necessary) conditions for
detectability in principle. So, for example  $\mu \;\Omega_{m0} +
\nu\;\Omega_{\Lambda0} >1\,$ is a sufficient condition for detectability
in principle of the topology of spherical universes ($\Omega_0 > 1$).

Before we proceed further it is worth stressing that the
conditions~(\ref{crite1}) and~(\ref{crite2}) are global (observer
independent), since they depend on $r_{inj}^{\:M}\,$, which is half
the length of the smallest geodesic in $M$. However, it is also possible
to define $r_{inj}^{\;}(x)$, half the length of the smallest geodesic
that passes through the point $x \in M$. Then~(\ref{crite1})
and~(\ref{crite2}) would become location (observer)-dependent conditions
for undetectability of cosmic topology.

\subsection{Secant method}

Consider the contour curve $\chi(z_{obs},\Omega_{m0},\Omega_{\Lambda0})=
r_{inj}^{\:M}\,$. For a given survey depth $z_{obs}$ we define the
secant line as the line joining the points $( \widetilde{\Omega}_{m0},0)$
and $(0,\widetilde{\Omega}_{\Lambda0})$ where the contour curve intersects
the axes $\Omega_{m0}$ and $\Omega_{\Lambda0}$, respectively. Clearly the
equation of this line is given by
\begin{equation}   \label{crit-extr}
\frac{\Omega_{m0}}{\widetilde{\Omega}_{m0}}+\frac{\Omega_{\Lambda0}}%
{\widetilde{\Omega}_{\Lambda0}}=1\;.
\end{equation}
For these intersecting points the redshift-distance relation%
~(\ref{chi(z)}) can also be more easily integrated. By writing explicitly
$\chi\,(z_{obs},\widetilde{\Omega}_{m0},0)= r_{inj}^{\:M}\,$ and
$\,\chi\,( z_{obs},0,\widetilde{\Omega}_{\Lambda0})= r_{inj}^{\:M}\,$,
and setting
$\varepsilon=\,[\mathrm{sign}(\,1 - \widetilde{\Omega}_{0})\,]^{\frac{1}{2}}$
we obtain, respectively%
\footnote{Incidentally, note that in these equations the functions
$\mathrm{arctanh}$ and $\log$ are defined as the analytical extensions
to the complex plane of the ordinary real functions, and as such they
apply to both spherical and hyperbolic cases.},

\parbox{14cm}{\begin{eqnarray*}
r_{inj}^{\:M} & = & 2\,\varepsilon \left[ \mathrm{arctanh}\,\,
(1- \widetilde{\Omega}_{m0})^{-1/2} - \,\mathrm{arctanh}\,\,
(\frac{1\,+\,z\:\widetilde{\Omega}_{m0}}{1\,-\,\widetilde{\Omega}_{m0}}
 \,)^{1/2} \right] \;,\nonumber \\
r_{inj}^{\:M} & =& \varepsilon\log \left\{\frac{ (1+z)\,
(1-\widetilde{\Omega}_{\Lambda0})^{1/2}\,+\,[\,(1+z)^{2}-z(2+z)
\:\widetilde{\Omega}_{\Lambda0}\,]^{1/2}}{1+
(1-\widetilde{\Omega}_{\Lambda0})^{1/2}}\:\right\} \;.
            \end{eqnarray*}}  \hfill
\parbox{1cm}{\begin{eqnarray}   \label{prov}  \end{eqnarray}}

We now wish to solve the equations~(\ref{prov}) for $\widetilde{\Omega}_{m0}$
and $\widetilde{\Omega}_{\Lambda0}\,$.
Although this can be done analytically, the (very long) resulting
expressions are not in general very useful, except in the limiting
case $z\rightarrow\infty.$ It can, however, always be done
numerically.

By treating $\Omega_{\Lambda0}$ as a function of $\Omega_{m0}$,
it is possible to show that for any fixed $z=z_{obs}$ the contour
curves $\chi(z_{obs},\Omega_{\Lambda0},\Omega_{m0})=$ $r_{inj}^{M}$
are convex (concave) for $\Omega_{0}<1\;$ ($\,>1\,$), i.e.,
$d^{2}\Omega_{\Lambda0}/d\Omega_{m0}^{2}\,>\,0\,$ ($<0\,$).
This property can also be gleaned from the parametric plot of
$\chi(\Omega_{\Lambda0},\Omega_{m0})\,$ (see for example%
~\cite{grt2001a}). It also ensures that the secant line crosses
the contour line only at the $\Omega_{m0}$ and $\Omega_{\Lambda0}$
axes, and that any sub-region of $\mathcal{U}$ lying between the
secant and the flat lines (region IV in Figure 2) will also lie
between the contour line and the flat line. Therefore the topology
of $M$ is undetectable for the values of the density parameters in
such region (IV). To sum up, if $\Omega_{m0}$ and
$\Omega_{\Lambda0}$ are such that inequality~(\ref{crite1}) holds
with $\alpha$ and $\beta$ given by~(\ref{crit-extr}), then the
topology of $M$ is undetectable.

Now we shall obtain a closed form expression for~(\ref{crit-extr}) in
the limiting case $ z\rightarrow\infty$.
Unless there is a horizon for a finite $z_{\max}$ (which is not the case
for currently accepted density parameters), $\chi(z)$ increases
monotonically with $z$. Therefore, by taking $z\rightarrow\infty$ we
obtain an upper bound for $\chi(z_{obs})$. Using~(\ref{prov}) we obtain
explicitly $\widetilde{\Omega}_{m0}$ and $\widetilde{\Omega}_{\Lambda0}$
in terms of $r_{inj}^{\:M}\,$ in this limiting case. The undetectability
conditions~(\ref{crite1}) [with (\ref{crit-extr})] then reduce to:

\bigskip

\textit{A universe with space section M has undetectable
topology if}

\parbox{14cm}{\begin{eqnarray*}
\cosh^{2} \,(\, \frac{r_{inj}^{\:M}}{2}\,)\; \Omega_{m0} +
\Omega_{\Lambda0} & >& 1\;, \quad \mathrm{for}
\;\quad \Omega_{0}<1\;,  \nonumber\\
\cos^{2}\,(\, \frac{r_{inj}^{\:M}}{2}\,)\; \Omega_{m0}+
\Omega_{\Lambda0} & < &1\;, \quad \mathrm{for}
\;\quad \Omega_{0}>1\;.
                \end{eqnarray*}}  \hfill
\parbox{1cm}{\begin{eqnarray}   \label{analit}  \end{eqnarray}}

Despite its simple form, this result is of considerable interest in that
it gives a test for undetectability for \emph{any} $z$. Note, however,
that for nearly flat universes the conditions~(\ref{analit}) and%
~(\ref{crite1}) for $z\sim1100$ (CMB) provide very close
results. Yet as these results are also close to those provided by the
tangent method (discussed in the next section), it follows that
for the cases of physical interest~(\ref{analit}) is, in practice, an
efficient criterion for undetectability of cosmic topology. Its closed
form greatly enhances its usefulness, as we shall illustrate
below.

A word of clarification is in order here. The expression used for
$\chi(z_{obs})$ here assumes only matter and cosmological constant
components.
For $z \gg 1100$ the photon energy density becomes significant. But
since its presence only decreases the actual value of $\chi(z_{obs})$
(increasing undetectability), the expressions~(\ref{analit}) remain
valid nonetheless.

It is worth stressing that the coefficients obtained by the secant line
method do not depend on the values of density parameters or their
associated uncertainties, and can be tabulated for any nearly flat
universe to be checked against present or future observations.

To close this subsection we note that there is a region in
the parameter plane (lying between the secant and contour
curves, depicted as region III in Figure 2) that does not
meet the condition~(\ref{crite1}), but for which the topology
of $M$ is still undetectable. In the next subsection we shall
present the tangent line method, and use it in
section~\ref{Examples} to show that this region is indeed
very narrow, and can be disregarded in practical cases of
physical interest (nearly flat universes).

\subsection{Tangent method}

The tangent line method discussed in more details in this section
provides necessary conditions for undetectability of cosmic topology
of nearly flat universes. As we have mentioned before their converses
also furnish conditions for detectability in principle of cosmic
topology of these universes.

We obtain a tangent to the contour line $\chi(z_{obs},\Omega_{m0},
\Omega_{\Lambda0}) =r_{inj}^{\:M}$ by taking a line passing through the
point $\bar{P}=(\bar{\Omega}_{m0},\bar{\Omega}_{\Lambda0})$ that is
perpendicular to the gradient of $\bigtriangledown\chi(z_{obs})$ at
$\bar{P}$. The equation for this tangent line can be written as
\begin{equation}  \label{tgline1}
\frac{\Omega_{\Lambda0}-\overline{\Omega}_{\Lambda0}}
{\Omega_{m0}-\overline{\Omega}_{m0}}=
-
\,\frac{\left.\frac{\partial\chi}{\partial\Omega_{m0}}\right|_{P=\bar{P}}}
{\left.\frac{\partial\chi}{\partial\Omega_{\Lambda0}}\right|_{P=\bar{P}}}\;.
\end{equation}
We note that the convexity (concavity) property mentioned above ensures
that this line will never cross the contour curve. Rearranging the terms
in~(\ref{tgline1}) we obtain
\begin{equation}  \label{coefs}
\frac{\frac{\partial\chi}{\partial\Omega_{m0}}}
{\frac{\partial\chi}{\partial\Omega_{m0}} \,\,\overline{\Omega}_{m0} +
\frac{\partial\chi}{\partial\Omega_{\Lambda0}}
\,\,\overline{\Omega}_{\Lambda0}} \;\Omega_{m0}
+\frac{\frac{\partial\chi}{\partial\Omega_{\Lambda0}}}{\frac{\partial\chi}
{\partial\Omega_{m0}}
\,\,\overline{\Omega}_{m0}+\frac{\partial\chi}{\partial\Omega_{\Lambda0}}
\,\,\overline{\Omega}_{\Lambda 0}}   \;\Omega_{\Lambda 0} =1\;,
\end{equation}
where (although we have not explicitly indicated for simplicity) all
partial derivatives are evaluated at $P=\bar{P}$.

The choice of $\bar{P}$ is not crucial, but in practice
we take it as the point in the contour curve that lies along the
gradient $\bigtriangledown \chi(z_{obs})$ from the point of maximum
$\chi(z_{obs})$ within the uncertainty region (see Fig.~\ref{SchemFig}).
The extreme points of uncertainty range given by WMAP data, together
with other measurements~\cite{WMAP}, are $P_{hyp}=(0.23,0.69)$ for the
hyperbolic, and $P_{sph}=(0.31,0.79)$ for the spherical case,
respectively.

In this way for a fixed $z=z_{obs}$ and for any given manifold $M$,
we obtain the tangent line at a point $\bar{P}$ of the contour
line $\chi(z_{obs})=r_{inj}^{M}\,$. Thus the necessary (but not
sufficient) conditions for the topology of nearly flat universes
to be undetectable are provided by~(\ref{crite2}) with $\mu$ and $\nu$
given by~(\ref{coefs}). Clearly if $\Omega_{m0}$ and $\Omega_{\Lambda0}$
are such that the converses of the inequalities~(\ref{crite2}) hold
with $\mu$ and $\nu$ given by~(\ref{coefs}), then the topology of the
universe is detectable in principle.

In terms of Figure 2, this means that for points in region I the
inequalities~(\ref{crite2}) hold, and therefore the topology
will be detectable in principle. There are also points (region II)
where the inequalities do not hold, and yet the topology is still
detectable in principle. This is however a very small region, as
we shall discuss below.

To close this section we note that for each pair
($r_{inj}^{:M},z_{obs}$) we can calculate numerically the
coefficients $\mu$ and $\nu$, and compare their values to $\alpha$
and $\beta\,$ obtained by the secant method to make clear that the
sub-region of $\,\mathcal{U}\,$ between the secant and the tangent
lines (regions II and III in Figure 2) is indeed very small for
manifolds whose contour curves intersect the uncertainty region.
This amounts to saying that in practice, for currently accepted
values of the density parameters, the secant line method as formulated
in eq.~(\ref{crite1}) [or its stronger closed form~(\ref{analit})]
gives an efficient criterion for undetectability of nearly flat
spherical and hyperbolic universes. In the next section we shall
discuss this point further.

\section{Applications}  \label{Examples}

In section~\ref{criteria} we described how to obtain the coefficients
($\alpha, \beta\,$) and ($\mu, \nu$) required by the secant and
tangent methods to ensure, respectively, the sufficient and necessary
conditions for undetectability [see~(\ref{crite1}) and~(\ref{crite2})],
providing therefore a systematic criterion for undetectability of
cosmic topology of nearly flat universes, given the unavoidable
observational uncertainties in the values of the density parameters.
We are mostly interested in undetectability conditions for large redshifts
($z=1100$ and $z \to \infty$), so here we shall tabulate these coefficients
for specific topologies  to provide support for the assertion that the
closed form expression~(\ref{analit}) is an accurate criterion for
undetectability of cosmic topology of nearly flat universes. For smaller
$z$, however, the coefficients must be calculated numerically.

The conditions~(\ref{crite1}) [with $\alpha$ and $\beta$
fixed by~(\ref{crit-extr})], their closed forms~(\ref{analit}), as well
as~(\ref{crite2}) [with the $\mu$ and $\nu$ given by~(\ref{coefs})] can
be trivially rewritten in terms of $\Omega_{0}$ and $\Omega_{X0}$,
where the latter variable can be either $\Omega_{\Lambda0}$ or
$\Omega_{m0}$.
These new variables make clearer for which exact values of the total
density parameter the topology of $M$ becomes undetectable. The
expression~(\ref{crite1}) then becomes
\begin{equation}
\begin{array}
[c]{c}%
\Omega_{0}>K+\gamma\,\Omega_{X0}\;, \quad \mathrm{for}
\quad \Omega_{0}<1\;, \\
\Omega_{0}<K+\gamma\,\Omega_{X0}\;, \quad \mathrm{for}
\quad \Omega_{0}>1\;.
\end{array}
\end{equation}

To make a direct  comparison with previous results we choose
$\Omega_{X0}$ as $\Omega_{\Lambda0}$ or $\Omega_{m0}$. In these
cases one has  $K=1/\alpha$ and $\gamma=(\alpha-\beta)/\alpha$ for
the pair ($\Omega _0,\Omega _{\Lambda}$), while for the pair
($\Omega _0,\Omega _{m0}$) one obtains $K=1/\beta\,$ and
$\,\gamma =(\beta-\alpha)/\beta$.
Now for the limit $z\rightarrow\infty$ the undetectability
conditions~(\ref{analit}) take the form:

\bigskip

\textit{A universe with space section\/}$\:M$\textit{\ has undetectable
topology if}

\parbox{14cm}{\begin{eqnarray*}
\Omega_{0} & > &\mathrm{sech^2}\,(\,r_{inj}^{M}\,/\,2\,)\,
+\,\tanh^{2}\,(\,r_{inj}^{M}\,/\,2\,)\,\; \Omega_{\,\Lambda0}\;, \quad
\mathrm{for} \quad \Omega_{0}<1\;, \nonumber \\ \Omega_{0} & < &
\sec^2\,(\, r_{inj}^{M}\,/\,2\,) \:\:\, - \:\, \tan^2\,(\,
r_{inj}^{M}\,/\,2\,)
\;\: \Omega_{\,\Lambda0}\;,\quad \:\, \mathrm{for}\quad \Omega_{0}>1 \;,
            \end{eqnarray*}}  \hfill
\parbox{1.5cm}{\begin{eqnarray}   \label{analit2} \end{eqnarray}}
or, in terms of $\Omega_{m0}$, if

\parbox{14cm}{\begin{eqnarray*}
\Omega_{0} & > & 1 \,
-\,\sinh^{2}\,(\,r_{inj}^{M}\,/\,2\,)\,\; \Omega_{m0}\;, \quad
\mathrm{for} \quad \Omega_{0}<1\;, \nonumber \\ \Omega_{0} & < &
1 \: + \:\, \sin^2\,(\,r_{inj}^{M}\,/\,2\,)
\;\: \Omega_{\,m0}\;,\quad \: \mathrm{for}\quad \Omega_{0}>1 \;.
            \end{eqnarray*}}  \hfill
\parbox{1cm}{\begin{eqnarray}   \label{analit3} \end{eqnarray}}
As an example, we follow~\cite{grt2001a} and tabulate $K$ and $\gamma$
for the first seven manifolds in the Hodgson-Weeks census of
hyperbolic manifolds, ordered here by decreasing $r_{inj}\,$.

\begin{table}[bht]
\centering
\par%
\begin{tabular}
[c]{|l|c|c|c|c|c|c|c|}\hline
{\small Manifold} & $r_{inj}$ & \multicolumn{2}{|c|}{{\small TL Method\/}
$z=1100$} & \multicolumn{2}{|c|}{{\small SL Method\/} $z=1100$} &
\multicolumn{2}{|c|}{{\small SL Method\/} $z\rightarrow\infty$}\\\cline{3-8}
& \hspace{1.3cm} & $\hspace{0.5cm}K\hspace{0.5cm}$ & $\gamma$ & $\hspace
{0.5cm}K\hspace{0.5cm}$ & $\gamma$ & $\hspace{0.5cm}K\hspace{0.5cm}$ &
$\gamma$\\ \hline \hline
$m007(3,1)$ & 0.416 & 0.955 & 0.040 & 0.956 & 0.045 & 0.958 & 0.041\\\hline
$m009(4,1)$ & 0.397 & 0.959 & 0.037 & 0.959 & 0.040 & 0.961 & 0.039\\\hline
$m003(-3,1)$& 0.292 & 0.977 & 0.020 & 0.978 & 0.022 & 0.978 & 0.022\\\hline
$m003(-2,3)$& 0.289 & 0.978 & 0.020 & 0.978 & 0.022 & 0.979 & 0.020\\\hline
$m003(-4,3)$& 0.288 & 0.978 & 0.020 & 0.978 & 0.022 & 0.980 & 0.020\\\hline
$m004(6,1)$ & 0.240 & 0.985 & 0.014 & 0.985 & 0.015 & 0.985 & 0.014\\\hline
$m004(1,2)$ & 0.183 & 0.991 & 0.008 & 0.991 & 0.009 & 0.991 & 0.008\\\hline
\end{tabular}
\caption{Coefficients $K=1/\alpha$ and $\gamma=(\alpha-\beta)/\alpha$
for undetectability of hyperbolic manifolds provided by the
necessary conditions from the tangent line (TL) method, and the
sufficient conditions given by the secant (SL) method, for two distinct
survey depths.}%
\label{Table1}%
\end{table}

Table~\ref{Table1} shows that both methods provide very close
numerical values for the coefficients $K$ and $\gamma$ when $z=1100$
and $z\rightarrow\infty$ (for the secant method).
This means that, on the one hand, the sufficient and
necessary undetectability conditions provided, respectively, by the
secant and tangent line methods are consistent (as expected), and
give very close results. On the other hand, this numerical
closeness also makes clear how narrow is the thin strip between
the secant and the tangent lines.

As an example, in~\cite{grt2001a} (see also~\cite{grt2002}) it was
shown that in the range $\Omega_{\Lambda0}\in[0.63,0.73]$ the
topology of both manifolds $m007(3,1)$ and  $m009(4,1)$ are
undetectable for $\Omega_{0} \in [0.99,1)$, and detectable in
principle for $\Omega_{0}\in[0.98,0.99)$.
Using the values in Table~\ref{Table1} from the conditions%
~(\ref{analit2}) we can state more precisely that the topology of
$m007(3,1)$ is undetectable for $\Omega_{\Lambda0}=0.63$ if
$\Omega_{0}>0.984$ and also for $\Omega_{\Lambda0}=0.73$ if
$\Omega_{0}>0.986$. Likewise, the topology of $m009(4,1)$ is
undetectable for $\Omega_{\Lambda0}=0.63$ if $\Omega_{0}>0.986$
and for $\Omega_{\Lambda0}=0.73$ if $\Omega_{0}>0.988$.

On the other hand, using the coefficients obtained with the
tangent method, it is clear that the topology of $m007(3,1)$
is detectable in principle for $\Omega_{0}<0.980$ if
$\Omega_{\Lambda0}=0.63$ and the topology of $m009(4,1)$ is
detectable in principle for $\Omega_{0}<0.982$ if
$\Omega_{\Lambda0}=0.73$.

These results agree with those obtained in~\cite{NEWweeks},
where it was found that the probability of detection of the
topology of these manifolds is zero, if $\Omega_{0}=0.99$.
The above examples also demonstrate clearly that the conditions
for (un)detectability of cosmic topology we have obtained
in this article extend previous results~\cite{grt2001a}~--%
~\cite{grt2002}, in which the undetectability of cosmic topology
was investigated only for specific values of the density parameters,
rather than values within the whole uncertainty region.

It is clear that $\chi_{obs}$  is not strongly dependent on
$\Omega_{\Lambda0}$, and it can be viewed as a function of
$\Omega_{0}$ for an appropriate value (fixed by observation,
for example) of $\Omega_{\Lambda0}$. Furthermore, as we have
seen for manifolds of physical interest the secant line with
$z\rightarrow\infty$ is a good approximation of the contour curve.
Therefore~(\ref{analit2}) can be understood as an analytical
relation between $\chi_{obs}$  and $\Omega_{0}$ which can be
used to answer specific questions about whole classes of
manifolds, as we shall discuss in examples below.

We now discuss some important examples related to spherical manifolds.
In Table~\ref{Table2} we tabulate all single action spherical manifolds
with their respective $r_{inj}$, as well as the lower and upper bounds
for the minimal value of $\Omega_{0}$ for undetectability provided,
respectively, by the secant line and the tangent line methods.
Single action manifolds are globally homogeneous and therefore
the (un)detectability conditions provided by both methods are location
independent, for any fixed survey depth. As a consequence, the tangent
line method also gives sufficient conditions for detectability (not
only in principle). For direct comparison with a similar table
in~\cite{NEWweeks2} we have taken $\Omega_{m0}=0.35$.
The minimal values of the total density in Table~\ref{Table2}
given by the tangent and secant line methods are in good agreement
with one another, as well as with those found in ref.~\cite{NEWweeks2}.

\begin{table}[htb] \centering
\begin{tabular}
[c]{|c|c|c|c|}\hline {\small Group} & $r_{inj}$ & {\small \ Sup.\/} $\Omega_{0}$&
{\small Inf.\/} $\Omega_{0}$\\ &  &{\small TL Method\/} $z=1100$ & SL{\small
\ Method\/} $z\rightarrow\infty $\\ \hline \hline
{\small Binary icosahedral\/} $I^{\ast}$ &
$\frac{\pi}{10}$& 1.011 & 1.009\\ \hline
{\small Binary octahedral\/} $O^{\ast}$&
$\frac{\pi}{8}$ & 1.018 & 1.013\\ \hline
{\small Binary tetrahedral\/} $T^{\ast}$&
$\frac{\pi}{6}$ & 1.031 & 1.023\\ \hline
{\small Binary dihedral\/} $D_{m}^{\ast}$&
$\frac{\pi}{2m}$& --- & $1+\sin^{2}\left(\frac{\pi }{4m}\right) \Omega _{m0\;}$
\\ \hline
{\small Cyclic } $Z_{n}$ & $\frac{\pi}{n}$& --- &$1+\sin^{2}\left(\frac{\pi}
{2n}\right)\Omega _{m0\;}$\\\hline
\end{tabular}
\caption{
Upper and lower bounds for the minimal value of $\Omega_0$ for undetectability
of topology, obtained, respectively, from the necessary conditions given by
the tangent line (TL) method, and the sufficient conditions provided by the
secant (SL) method, for $\Omega_{m0}=0.35$.
 \label{Table2}  }
\end{table}

It is useful to employ the analytical expression~(\ref{analit3}) to
provide a lower bound for undetectability in the case of binary
dihedral and cyclic spaces, since it contains an explicit dependence
on $\Omega_{m0}$, allowing a more systematic study of detectability.
For these manifolds the undetectability conditions are given as
functions of $n$ and $m$, respectively. Up to second order they
reduce to $1+ 0.86/(2m)^{2}$ and $1+0.86/n^{2}$, for
$\Omega_{m0}=0.35$, in agreement with table~2 in~\cite{NEWweeks2}.

\begin{table}[htb] \centering
\begin{tabular}
[c]{|c|c|c|c|}\hline
$m$ & $n$ & $r_{inj}$ & {\small Max.\/} $\Omega_{0}$ {\small \ }\\
&  &  & {\small TL Method\/} $z=1100$ \\ \hline \hline
2 & 4 & 0.785 & 1.072\\\hline
3 & 6 & 0.524 & 1.031\\\hline
4 & 8 & 0.393 & 1.018\\\hline
5 & 10 & 0.314 & 1.011\\\hline
6 & 12 & 0.262 & 1.008\\\hline
9 & 18 & 0.175 & 1.005\\\hline
\end{tabular}
\caption{
Upper bound for the minimal value for $\Omega_0$ for undetectability of
cosmic topology obtained from the (necessary) conditions provided by
the tangent line method for sample binary dihedral $D^\ast_m$ and
cyclic spaces $Z_n$, for $\Omega_{m0}=0.35$. \label{Table3} }
\end{table}%

The values in Tables~\ref{Table2} and~\ref{Table3} show that the
results given by the tangent and secant line methods are in good
agreement with each other, as well as with those in~\cite{NEWweeks2}.
For manifolds with smaller $r_{inj}$ the results of course become
closer.

The values in Tables~\ref{Table2} and~\ref{Table3} show
that the results given by the tangent and secant line methods are
in good agreement with each other, as well as with those
in~\cite{NEWweeks2}. If we compare the bounds in Table
\ref{Table2} with numerical estimates of the contour curves with
$\Omega_{m0}=0.35$, it becomes clear that the actual
minimal value of $\Omega_{0}$ for detectability is very
close (agreeing to 3 decimal places) to the bound given by the
tangent method. This is not surprising, given that the tangent
line is the best linear fit of the contour curve. We note, however,
that the use of the secant line instead  does not introduce
significant errors.

To complete the picture we also show in Table~\ref{Table3} the maximum
values of $\Omega_{0}$ for detectability provided by the tangent method
for some members of the $D_{m}^{\ast}$ and $Z_{n}$ classes, again with
$\Omega_{m0}=0.35$.
These two classes are particularly important because as
$\Omega_{0}\rightarrow 1$ from above, there is always a $n_{\ast}$
(or $m_{\ast}$) such that the topology corresponding to $Z_{n}$
(or $D_{m}^{\ast}$) is detectable for $n~>~n_{\ast}$ (or $m > m_{\ast}$).
In particular, the single action cyclic spaces are globally homogeneous
and constitute a subclass of the lens spaces $L(n,q)$, namely $L(n,1)$.
The general lens spaces $L(n,q)$ for $q \neq 1$ are not globally
homogeneous, but have the same injectivity radius, $r_{inj}=\pi/n$, of
the homogeneous family. Thus, although for the homogeneous cyclic spaces
the condition $n>n_{\ast}$ is global and ensures that the topology is
detectable, for the inhomogeneous cases it becomes a condition for
detectability in principle only. For some specific density values
$n_{\ast}$ has been obtained~\cite{grt2001a}. Here we extend these
results by solving~(\ref{analit3}) for $n_{\ast}$ and $m_{\ast}$ to
obtain

\parbox{14cm}{\begin{eqnarray*}
n_{\ast} & = &\mathrm{Int}\left\{\,\frac{\pi}{2}\,\left[\,\arcsin\sqrt{\,
\frac{\Omega_{0}-1}
 {\Omega_{m0}}\;}\:\,\right]^{-1} \:\right\} \;, \nonumber \\
 m_{\ast} & = &\mathrm{Int}\left\{\,\frac{\pi}{4}\,\left[\,\arcsin\sqrt{\,
\frac{\Omega_{0}-1}
 {\Omega_{m0}}\;}\:\,\right]^{-1}\:\right\} \;,
 \end{eqnarray*} }  \hfill
\parbox{1cm}{\begin{eqnarray} \label{m*} \end{eqnarray} }
where $\mathrm{Int}[x]$ denotes the integer part of $x$.

These expressions can be used to calculate $n_{\ast}$ and $m_{\ast}$\
for any set of density parameters. For $z\rightarrow\infty$ expressions%
~(\ref{m*}) are in agreement with results obtained~\cite{grt2001a},
which in turn are very close to the results for $z=3000$ tabulated
in~\cite{grt2001a}.
For smaller $z$ the values of $n_{\ast}$ and $m_{\ast}$ must be
obtained numerically, by using equation~(\ref{prov}).
We stress that these expressions also provide an example of how
our results can be used to study the detectability of cosmic
topology for classes of manifolds systematically without
resorting to numerical calculations of the contour lines.

 \section{Conclusions and final remarks}
\label{Conclusion}

Current estimates of the cosmological density parameters suggest
that the universe is nearly flat, with the associated $2\sigma$
confidence level allowing for the possibility of a spherical,
flat or hyperbolic universe. Motivated by this we have
reexamined the question of detectability of the cosmic topology
taking into account the uncertainty region in the
$\Omega_{\Lambda}$~--~$\Omega_{m}$ parametric plane. Since the
detectability or undetectability of the cosmic topology crucially
depends on the values of the density parameters, the true
determination of the topology requires taking into account a
detailed analysis of the effects of the related
uncertainties.

We present two complementary methods (secant and tangent methods)
that give, respectively, sufficient (but not necessary) conditions for
undetectability, and necessary (but not sufficient) conditions for
undetectability of cosmic topology. The converses of the latter also
constitute a set of sufficient (but not necessary) conditions for
detectability in principle. These methods are systematic in the
sense that they determine (un)detectability for any values of the
density parameters in the uncertainty region, except for a negligible
thin strip.

When applied to specific manifolds, both methods provide conditions
that are shown to be in good agreement, for any fixed survey depth,
accurately separating the parameter plane into undetectable and
detectable regions. Since they are both systematic and accurate,
these criteria extend previous results~\cite{grt2001a}~--~\cite{grt2002},
(see also~\cite{NEWweeks}) in which the undetectability of cosmic topology
was investigated only for specific values of the density parameters, rather
than within an uncertainty region that include such values.

Clearly the tangent line obtained from~(\ref{coefs}),
being the best linear fit for the contour curve, is a better
approximation of the contour curve than the secant
line~(\ref{crit-extr}). However, as was discussed in
Sections~\ref{criteria} and~\ref{Examples}, in practice the
difference between the coefficients given by the tangent
and secant line methods is small, and become even smaller
as $\Omega_{0}\rightarrow1$, and which one is used is a
matter of convenience. The closed form of the latter in the
limit $z\rightarrow\infty$ makes it more useful to study
classes of manifolds rather than specific examples.

An important result of this article is the closed form
conditions~(\ref{analit}) for undetectability of cosmic topology or
nearly flat universes obtained from the secant method in the
$z\rightarrow\infty$ limit. These inequations  can be
seen, to a very good approximation, as establishing conditions for
detectability in principle as well, as can be shown by comparison
with numerical values obtained from both methods for $z=1100$.
For high redshifts we can therefore use~(\ref{analit}) to separate
the parameter plane into undetectable and detectable sub-regions with
great accuracy. The closed form of these conditions makes its
application quite straightforward and potentially more useful.
Equation~(\ref{m*}) is a good example of such application,
because it extends previous results to a more general, and yet
simpler, form.

There are other advantages in the use of~(\ref{analit}) and its
counterparts~(\ref{analit2}) and~(\ref{analit3}). If, as
expected, new observations further constrain the uncertainty
region nearer to the flat line, the conditions provided by the
secant and tangent methods will become closer to each other,
and either~(\ref{analit}), (\ref{analit2}) or (\ref{analit3})
will become even more accurate. In that case, our expressions
can be used to address a number of questions regarding the
detectability of cosmic topology for classes of manifolds, as
for example in the detectability conditions $n>n^{\ast}$ and
$m>m^{\ast}$ with $n^{\ast}$ and $m^{\ast}$ given by (\ref{m*}).
Note also that, for any given survey depth, the coefficients in
the conditions~(\ref{analit}) or its counterparts for each (fixed)
manifold do not depend on the size and shape of the uncertainty
region.

In a recent paper Weeks~\cite{NEWweeks} introduced the so called
injectivity profile, in which for each manifold the probability of
detecting the cosmic topology is plotted as a function of the horizon
distance $\chi(z_{hor})$. The probability is defined as the fraction
of the manifold where
$\chi(z_{obs})\leq r_{inj}^{\;}(x) \leq\chi(z_{obs})+\Delta\chi$.
We note that even though the (un)detectability conditions presented
here were obtained for a global $r_{inj}^{M}$, expressions such
as~(\ref{analit}) still hold if the injectivity radius function
$r_{inj}^{\;}(x)$ is used instead.
Such location dependent conditions could be used to calculate
injectivity profiles in terms of $\Omega_{0}$ instead of $\chi(z_{hor})$.
This would allow us to determine, for instance, how the probability of
detectability changes as the total density approaches its critical
value.

Finally even though we have used a $\Lambda$CDM framework, similar
methods could be developed for other cosmological models with different
redshift-distance relations in order to obtain conditions for undetectabilty of
cosmic topology.

\section*{Acknowledgments}

We thank CNPq, MCT/CBPF and FAPERJ for the grants under which this work
was carried out. We also thank G.I. Gomero and A.F.F. Teixeira for fruitful
discussions.

\end{document}